# A Note on Implementing a Special Case of the LEAR Covariance Model in Standard Software

Sean L. Simpson[†]

*Wake Forest School of Medicine, Winston-Salem, NC, USA*

Min Zhu

*SAS Institute Inc., Cary, NC, USA*

Keith E. Muller

*University of Florida, Gainesville, FL, USA*

[†]Correspondence to: Sean L. Simpson, Department of Biostatistical Sciences, Wake Forest School of Medicine, Winston-Salem, NC 27157-1063

**ABSTRACT**

Repeated measures analyses require proper choice of the correlation model to ensure accurate inference and optimal efficiency. The *linear exponent autoregressive* (LEAR) correlation model provides a flexible two-parameter correlation structure that accommodates a variety of data types in which the correlation within-sampling unit decreases exponentially in time or space. The LEAR model subsumes three classic temporal correlation structures, namely compound symmetry, continuous-time AR(1), and MA(1), while maintaining parsimony and providing appealing statistical and computational properties. It also supplies a plausible correlation structure for power analyses across many experimental designs. However, no commonly used statistical packages provide a straightforward way to implement the model, limiting its use to those with the appropriate programming skills. Here we present a reparameterization of the LEAR model that allows easily implementing it in standard software for the special case of data with equally spaced temporal or spatial intervals.

**1. Introduction**

Repeated measures designs are ubiquitous in scientific research, often being employed to examine longitudinal, spatial, or spatio-temporal data. Appropriate analysis requires proper specification of the correlation pattern induced by these experimental designs. For example, Muller et al. (2007) and Gurka et al. (2011) showed that under-specifying the correlation structure can severely inflate test size of tests of fixed effects in the general linear mixed model. Thus valid inference requires enough flexibility of the correlation model to allow reasonable fidelity to the true pattern. On the other hand, optimal efficiency and the existence of computable estimates demand a parsimonious parameterization of the correlation pattern. The *linear exponent autoregressive* (LEAR) correlation model is a flexible two-parameter structure that satisfies these two opposing goals for situations in which the within-sampling unit correlation decreases exponentially in time or space (Simpson et al., 2010). It allows for an attenuation or acceleration of the exponential decay rate imposed by the commonly used continuous-time

AR(1) structure (also referred to as the exponential model). The LEAR model and its derivatives are useful in a variety of contexts for data analysis (Simpson et al., 2010; Simpson 2010; Simpson and Edwards, 2013; Simpson et al., 2014a; Simpson et al., 2014b). The LEAR model also provides powerful tools for power analyses as noted in the documentation for GLMPOWER (SAS Institute, 2013), GLIMMPSE (Kreidler et al., 2013, http://glimmpse.SampleSizeShop.org/), and POWERLIB (Johnson et al., 2009). However, no commonly used statistical package provides a straightforward way to implement the model, limiting its use to those with the appropriate programming skills. To address this barrier for the important special case of data with equally spaced temporal or spatial intervals, we present a reparameterization of the LEAR model that allows easily implementing it in standard software.

## 2. LEAR Reparameterization and Software Implementation

*2.1 LEAR Definition*

Suppose $\boldsymbol{y}_i$ is a $p_i \times 1$ vector of $p_i$ observations on the $i^{th}$ independent sampling unit (ISU; "subject") $i \in \{1, \ldots, N\}$, with corresponding measurement times or locations $\boldsymbol{t}_i$. A corresponding distance between measurement times or locations is $d_{ijk} = d(t_{ij}, t_{ik})$, with extremes of $d_{\min} = \min_{ijk}\{d_{ijk}\}$ and $d_{\max} = \max_{ijk}\{d_{ijk}\}$. With $\mathcal{C}(\,\cdot\,)$ the correlation operator, $\rho_{i;jk} = \mathcal{C}(y_{ij}, y_{ik})$ indicates the correlation between a pair observations within an ISU. The *linear exponent autoregressive* (LEAR) correlation structure (Simpson et al., 2010) has correlation matrix $\boldsymbol{\Gamma}_i = \{\rho_{i;jk}\}$ with

$$\rho_{i;jk} = \mathcal{C}(y_{ij}, y_{ik}) = \begin{cases} \rho_L^{d_{\min} + \delta[(d_{ijk} - d_{\min})/(d_{\max} - d_{\min})]} & j \neq k \\ 1 & j = k \end{cases}, \quad (1)$$

$0 \leq \rho_L < 1$ and $0 \leq \delta$.

When $\delta = 0$, the model reduces to the well known equal correlation model for which the correlation between measurements on a given sampling unit is fixed at $\rho_L$ no matter how far apart in time or space the measurements were taken. When $0 < \delta < d_{\max} - d_{\min}$, the correlation between measurements on a given sampling unit decreases in time or space at a slower rate than that imposed by the AR(1) model. When $\delta = d_{\max} - d_{\min}$, the model reduces to the AR(1) correlation model. When $\delta > d_{\max} - d_{\min}$, the correlation between measurements on a given sampling unit decreases in time or space at a faster rate than that imposed by the AR(1) model. As $\delta \to \infty$, this model approaches the moving average model of order 1, MA(1). Though values of $\delta < 0$ yield valid autocorrelation functions for which the correlation between measurements on a given sampling unit would increase with increasing time or distance between measurements, such patterns are rare; therefore the parameter space is restricted for reasons of practicality. If we assume an equal variance structure then

$$\mathcal{V}(y_{ij}, y_{ik}) = \sigma^2 \begin{cases} \rho^{d_{\min} + \delta[(d_{ijk} - d_{\min})/(d_{\max} - d_{\min})]} & j \neq k \\ 1 & j = k, \end{cases} \quad (2)$$

where $\mathcal{V}(\,\cdot\,)$ is the covariance operator.

*2.2 Reparameterization and Software Implementation*

A useful special case of the LEAR model meets two restrictions: 1) all sampling units have equally spaced integer values of $\{d_{ijk}\}$ and 2) $d_{\min} = 1$. The special case allows reparameterizing the LEAR model as an ARMA(1, 1) model as detailed here. If $\delta_d = \delta/(d_{\max} - d_{\min})$, then

$$\rho_L^{d_{\min} + \delta[(d_{ijk} - d_{\min})/(d_{\max} - d_{\min})]} = \rho_L^{d_{\min}} \rho_L^{\delta_d(d_{ijk} - d_{\min})}. \quad (3)$$

In turn

$$\rho_L^{d_{\min}} \rho_L^{\delta_d(d_{ijk} - d_{\min})} = \rho_L \rho_L^{\delta_d(d_{ijk} - 1)}. \quad (4)$$

Equally spaced integer values for all sampling units implies $d_{ijk} = |j - k|$. If $\tau = \rho_L$ and $\rho_A = \rho_L^{\delta_d}$, then

$$\rho_L \rho_L^{\delta_d(d_{ijk} - 1)} = \tau \rho_A^{|j - k| - 1}, \quad (5)$$

which is the correlation function component of the ARMA(1,1) model. Finally, adding the variance parameter to the reparameterization gives

$$\mathcal{V}(y_{ij}, y_{ik}) = \sigma^2 \begin{cases} \tau \rho_A^{|j-k|-1} & j \neq k \\ 1 & j = k, \end{cases} \quad (6)$$

which defines the ARMA(1,1) covariance structure. Thus, for data with equally spaced temporal or spatial intervals, the LEAR covariance model can be fit in standard software using the procedures noted in Table 1.

Table 1. Standard Statistical Software with an ARMA(1,1) Covariance Option

| Software | Procedures/Packages |
|----------|---------------------|
| SAS      | Glimmix             |
|          | Mixed               |
| R        | nlme                |
| SPSS     | Mixed               |

## 3. Discussion

A reparameterization of the LEAR correlation model allows implementing it in standard software for the special case of equally spaced temporal or spatial intervals. The reparameterization provides a powerful tool for repeated measures data and power analysis. It is important to note that while the LEAR and ARMA(1,1) parameterizations are mathematically equivalent for the special case, there is no guarantee that the two parameterizations will lead to the same estimated covariance matrix given that the computability of estimates can vary greatly with the parameterization for nonlinear models such as the LEAR and ARMA(1,1).

**Acknowledgements**

Simpson's work was supported in part by NIBIB K25 EB012236-01A1. Muller's work was supported in part by NIGMS 1R25GM111901-01, NLM 1G13LM011879-0, and NIGMS 9R01GM121081-05. The authors thank John Castelloe from SAS Institute for his insightful comments and suggested edits that greatly improved the presentation of material.

# References

Gurka, M. J., Muller, K. E., and Edwards, L. J. (2011), "Avoiding Bias in Mixed Model Inference for Fixed Effects," *Statistics in Medicine*, 30, 2696–2707.

Johnson, J. L., Muller, K. E., Slaughter, J. C., Gurka, M. J., Gribbin, M. J., and Simpson, S. L. (2009), "POWERLIB: SAS/IML Software for Computing Power in Multivariate Linear Models," *Journal of Statistical Software*, 30, 1-27.

Kreidler, S. M., Muller, K. E., Grunwald, G. K., Ringham, B. M., Coker-Dukowitz, Z. T., Sakhadeo, U. R., Barón, A. E. and Glueck, D.H. (2013), "GLIMMPSE: Online Power Computation for Linear Models With and Without a Baseline Covariate," *Journal of statistical software*, 54(10), 1-27.

Muller, K.E., Edwards, L. J., Simpson, S. L., and Taylor, D.J. (2007), "Statistical Tests With Accurate Size and Power for Balanced Linear Mixed Models," *Statistics in Medicine*, 26, 3639-3660.

SAS Institute. (2013), *SAS, Version 9.4*. SAS Institute, Inc.: Cary, NC.

Simpson, S. L. (2010), "An Adjusted Likelihood Ratio Test for Separability in Unbalanced Multivariate Repeated Measures Data," *Statistical Methodology*, 7, 511-519.

Simpson, S. L., and Edwards, L. J. (2013), "A Circular LEAR Correlation Structure for Cyclical Longitudinal Data," *Statistical Methods in Medical Research*, 22, 296-306.

Simpson, S. L., Edwards, L. J., Muller, K. E., Sen, P. K., and Styner, M. A. (2010), "A Linear Exponent AR(1) Family of Correlation Structures," *Statistics in Medicine*, 29, 1825-1838.

Simpson, S. L., Edwards, L. J., Muller, K. E., and Styner, M. A. (2014a), "Kronecker Product Linear Exponent AR(1) Correlation Structures for Multivariate Repeated Measures Data," *PLoS One*, 9(2), e88864.

Simpson, S. L., Edwards, L. J., Muller, K. E., and Styner, M. A. (2014b), "Separability Tests for High-Dimensional, Low Sample Size Multivariate Repeated Measures Data," *Journal of Applied Statistics*, 41, 2450-2461.